\documentclass[
twocolumn,
showpacs,preprintnumbers,
amsmath,amssymb,
aps,
prl,
]{revtex4-1} 

\usepackage{graphicx,color}
\usepackage{dcolumn}
\usepackage{bm}
\usepackage{epstopdf} 

\begin{document}
\title{Probing water structures in nanopores using tunneling currents}
\author{P. Boynton and M. Di Ventra}
\affiliation{Department of Physics, University of California, San Diego, La Jolla, CA 92093-0319}
\date{\today}
%
\begin{abstract}
We study the effect of volumetric constraints on the structure and electronic transport properties of distilled water in a nanopore with embedded electrodes. Combining classical molecular dynamics simulations with quantum scattering theory, we show that the structural motifs water assumes inside the pore can be probed directly by tunneling. In particular, we show that the current does not follow a simple exponential curve at a critical pore diameter of about 8 {\AA}, rather it is larger than the one expected from simple tunneling through a barrier. This is due to a structural transition from bulk-like to ``nanodroplet'' water domains. Our results can be tested with present experimental capabilities to develop our understanding of water as a complex medium at nanometer length scales.
\end{abstract}
\maketitle

Liquid water is a very common and abundant substance that is considered a fundamental ingredient for life more than any other. Yet we do not fully understand many of its properties, especially when we probe it at the nanometer scale, although a lot of research has been done on this important system in this regime \cite{Hahn1998, Hugelmann2003L643, Kolesnikov2004, PhysRevE.85.031501, Hummer2001, PhysRevB.76.195403}. Some of the first experimental studies of water on the nanoscale have been done using a scanning-tunneling microscope (STM) \cite{Hahn1998}, in which the tunneling barrier height was found to be unusually low. This was hypothesized to be the result of the three-dimensional nature of electron tunneling in water. Some STM experiments actually studied the tunneling current as a function of distance to understand the solid/liquid interface and found that the tunneling current oscillates with a period that agrees with the effective spacing of the Helmholtz layers \cite{Hugelmann2003L643}. Water has also been studied when encapsulated by single-walled carbon nanotubes in which, via neutron scattering, the water was observed to form a cylindrical ``square-ice sheet'' which enclosed a more freely moving chain of molecules \cite{Kolesnikov2004}. These structures are related to the fact that these carbon nanotubes have cylindrical symmetry and are hydrophobic. More recently, the dynamics of water confined by hydrophilic surfaces were studied by means of inelastic X-ray scattering showing a phase change at a surface separation of 6 {\AA}. Well above 6 {\AA} there are two deformed surface layers that sandwich a layer of bulk-like water but below 6 {\AA} the two surface layers combine into one layer that switches between a localized ``frozen'' structure and a delocalized ``melted'' structure \cite{PhysRevE.85.031501}. On the computational side, many molecular dynamics (MD) simulations have been done to study water in a variety of environments. Of particular interest has been the study of hydrophobic channels because in this case water has been shown to escape from the channel altogether for entropic gain \cite{Hummer2001, PhysRevB.76.195403}. However, these structures, and in particular the formation of water nanodroplets, are difficult to probe experimentally.

Recent interest in fast DNA sequencing approaches has been crucial to the advancement of novel techniques to probe polymers in water environments at the nanometer scale. In particular, the proposal to sequence DNA by tunneling \cite{Lagerqvist2006, Krems2009} has been instrumental for the development of sub-nanometer electrodes embedded into nanochannels \cite{Tsutsui2008, Albrecht2012, Healy2012}. These techniques open the door to investigating the properties of liquids volumetrically constrained by several materials by relating the local structure of the liquid to electrical (tunneling) currents.

In this Letter, we take advantage of these newly-developed experimental techniques and propose the study of water in nanopores with embedded electrodes. We find that indeed the structural motifs water assumes inside pores of different diameters can be probed directly by tunneling. In fact, we predict that the tunneling current does not follow a simple exponential curve at a critical pore diameter of about 8 {\AA} as simple tunneling through a barrier would produce. Instead, water domains form a specific density of states which in turn gives rise to these peculiar features. Our findings can be tested with the available experimental capabilities on similar systems \cite{Tsutsui2008, Albrecht2012, Healy2012}.

To better understand the nature of this substance on the nanoscale, we study the effects of confinement on water's structure and electronic transport properties in silicon nitride nanopores using classical molecular dynamics (MD) combined with quantum transport calculations. Since the system is at room temperature quantum effects related to protons are negligible, which allows us to use NAMD 2.7 \cite{Phillips2005}, a highly parallel classical MD application. We have chosen to work with $\mbox{Si}_3\mbox{N}_4$ nanopores because they are readily fabricated to have very small constrictions. Note also that the environment we consider is not hydrophobic because silicon nitride ($\mbox{Si}_3\mbox{N}_4$) nanostructures are known to have dangling atoms that produce polar surfaces \cite{Zhang1992}.
\begin{figure}
\includegraphics[width=\columnwidth]{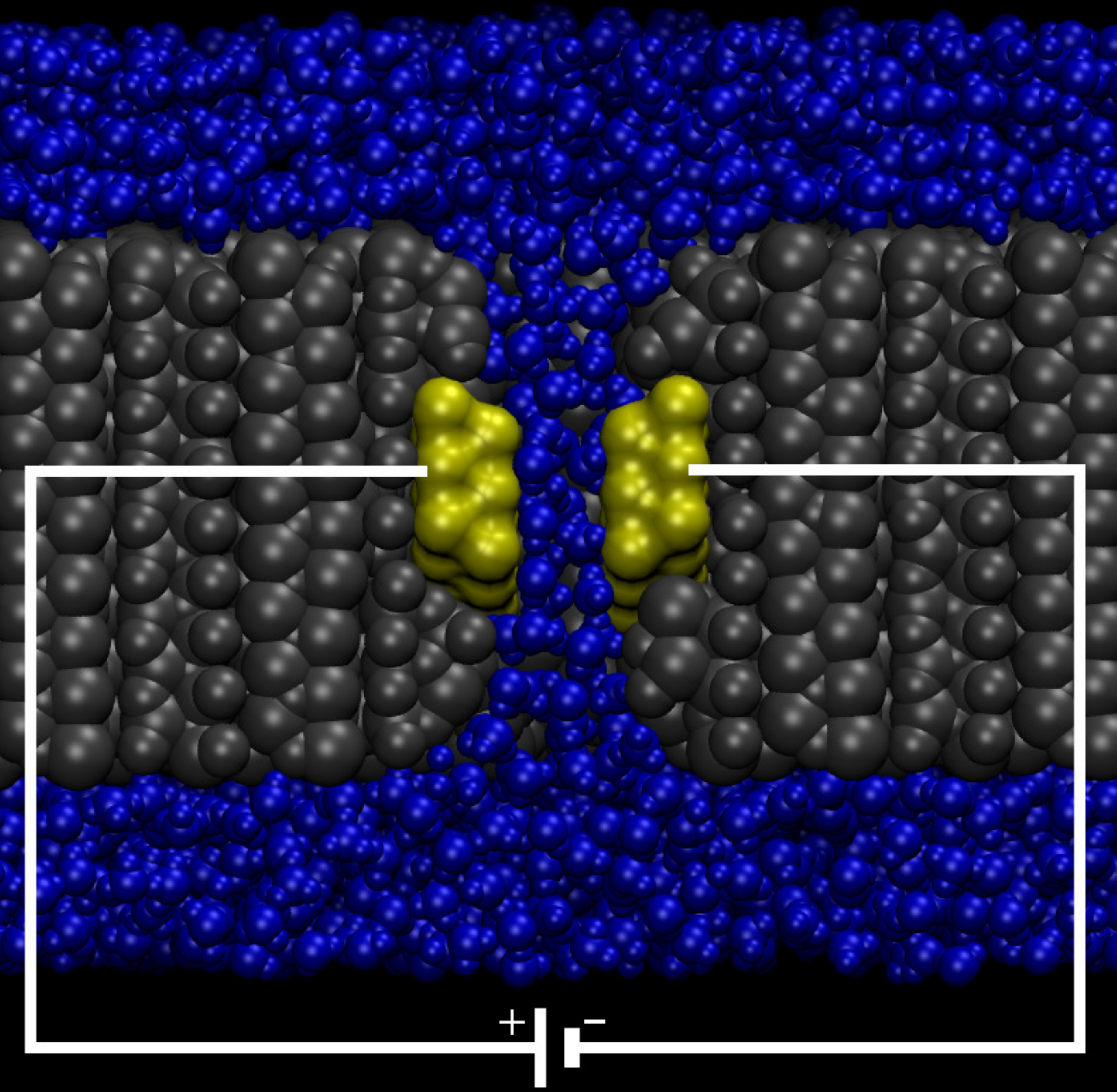}
\caption{\label{fig:setup}(Color online) A cross section of a nanopore for MD after equilibration. The $\mbox{Si}_3\mbox{N}_4$ (gray) membrane is cut into a hexagonal prism 2.6 nm thick and 9.1 nm wide (from one vertex to the opposing hexagonal vertex) to satisfy periodic boundary conditions in all three dimensions of space. We use a double-conical pore with a $10\,^{\circ}$ slant off of cylindrical to better represent currently fabricated nanopores. The $\mbox{Si}_3\mbox{N}_4$ is harmonically constrained to reproduce its dielectric behavior in experiments. The system is solvated to create a water (blue) reservoir above and below the pore of combined thickness 2.6 nm. The electric field goes from left to right between the gold (yellow) electrodes.}
\end{figure}

The system is built in a manner similar to the synthetic pore in \cite{Aksimentiev2011}. Using VMD \cite{HUMP96}, we build a $\beta\mbox{-Si}_3\mbox{N}_4$ membrane containing a double-conical pore with inner diameter ranging from 4.5 to 9.25 {\AA} (atom center to atom center). The membrane includes two fixed embedded gold electrodes that span the small constriction at the center of the pore, similar to the pairs of electrodes introduced in \cite{Lagerqvist2006, Krems2009}. Above and below the membrane lie water reservoirs of about 3200 molecules combined that provide a buffer between periodic images, and provide the bulk with which the water molecules in the pore can be recycled (see Fig. \ref{fig:setup}). We can safely ignore any entrance effects on the structure of the confined water between the electrodes since the correlation length of bulk water at room temperature is between 1.5 and 2 {\AA} \cite{Moore2009} and each pore entrance lies approximately 8 {\AA} from the region of interest. In addition, the confinement is gradual due to the double-conical shape and testing has been done on similar nanopores filled with water and ions to show that increasing the thickness of the pore beyond 2.6 nm does not change the basic dynamics of the system \cite{Krems2010}. We utilize the CHARMM27 force field \cite{Foloppe2000, MacKerell2000} for the interactions of the TIP3P water whereas we use UFF parameters for the $\mbox{Si}_3\mbox{N}_4$ \cite{Wendel1992}. Furthermore no ions are added to the system to permit the study of pure water structures and their effect on tunneling current. A Langevin thermostat keeps temperature set to 295 K with a damping coefficient of 1 $\mbox{ps}^{-1}$ applied to the $\mbox{Si}_3\mbox{N}_4$ while a bias of 1 V between the embedded electrodes is achieved using the Grid-steered Molecular Dynamics feature in NAMD 2.7 \cite{Phillips2005}. More specifically, we impose a 3-dimensional potential grid on the region between the gold electrodes which is linear in the transverse axis and constant in the remaining two. Padding must be added to the box so that there are no discontinuities in the field at the edges, after which the grid is interpolated by cubic polynomials. From these polynomials the gradient is taken to obtain the electric field, which has been checked to be accurate in the affected region. Note that we do not treat the image forces created by the polarized water's proximity to an equipotential surface (both electrodes). However, our calculations show that the strength of these forces is less than the force due to the 1 V bias, although within an order of magnitude for the water molecules on the electrode surface. Therefore, we expect that since the image forces act to align the water, much like the external field does, the structural effects we find can only be enhanced by their inclusion. The entire equilibrated system evolves over 5 ns in an NVT ensemble with 1 fs time steps, yet atomic coordinates are recorded every ps.

Position data snapshots of the gold and the surrounding water molecules are then taken to evaluate the current over time. Although each snapshot is only a static representation of the system we can safely calculate the tunneling current at each recording. This is because the time scale governing the tunneling electrons (${\sim}10^{-15}$ s) is much smaller than the time scale of the relevant dynamics of the water molecules (${\sim}10^{-12}$ s) \cite{Mitra2001}. The effect of dephasing and other inelastic effects have been estimated to be small for the distribution of currents at reasonable electron-molecular vibrations and rotational time scales \cite{Krems2009}, thus they can be neglected. To calculate the current we use a single-particle scattering approach that involves obtaining a tight-binding Hamiltonian (see, e.g., \cite{DiVentra2008}) and the single-particle retarded Green's function, as detailed in \cite{Krems2009} with no added noise. The resultant current is given by $I = \frac{2 e}{h} \int^{\infty}_{-\infty}{dE\,T(E)[f_r(E) - f_l(E)]}$ where $e$ is the elementary charge, $h$ is Planck's constant, $E$ is the energy of the scattering electron, $T$ is the total transmission function, and $f_r$ and $f_l$ are the right and left electrode Fermi-Dirac distribution functions, respectively.

\begin{figure}
\includegraphics[width=\columnwidth]{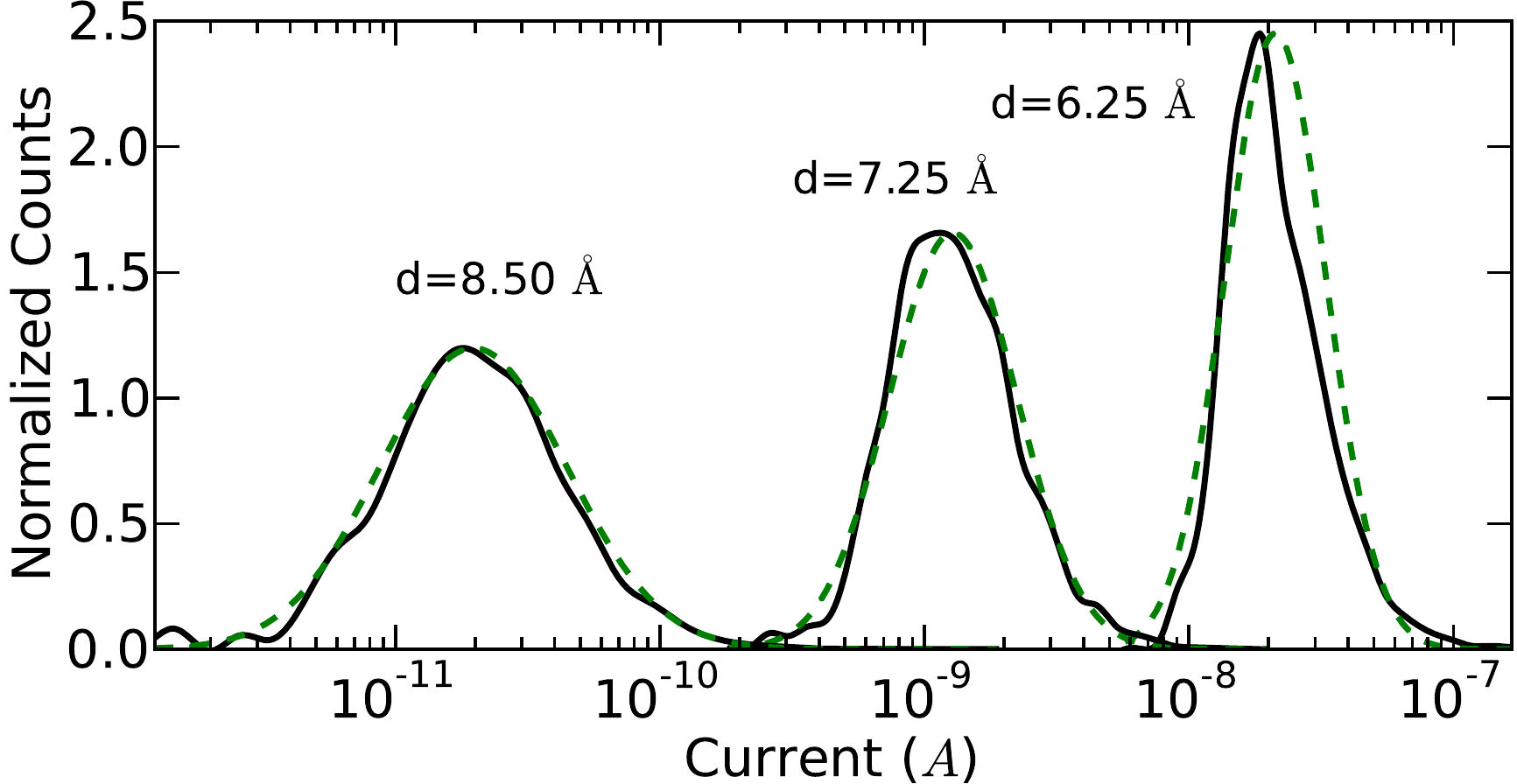}
\includegraphics[width=\columnwidth]{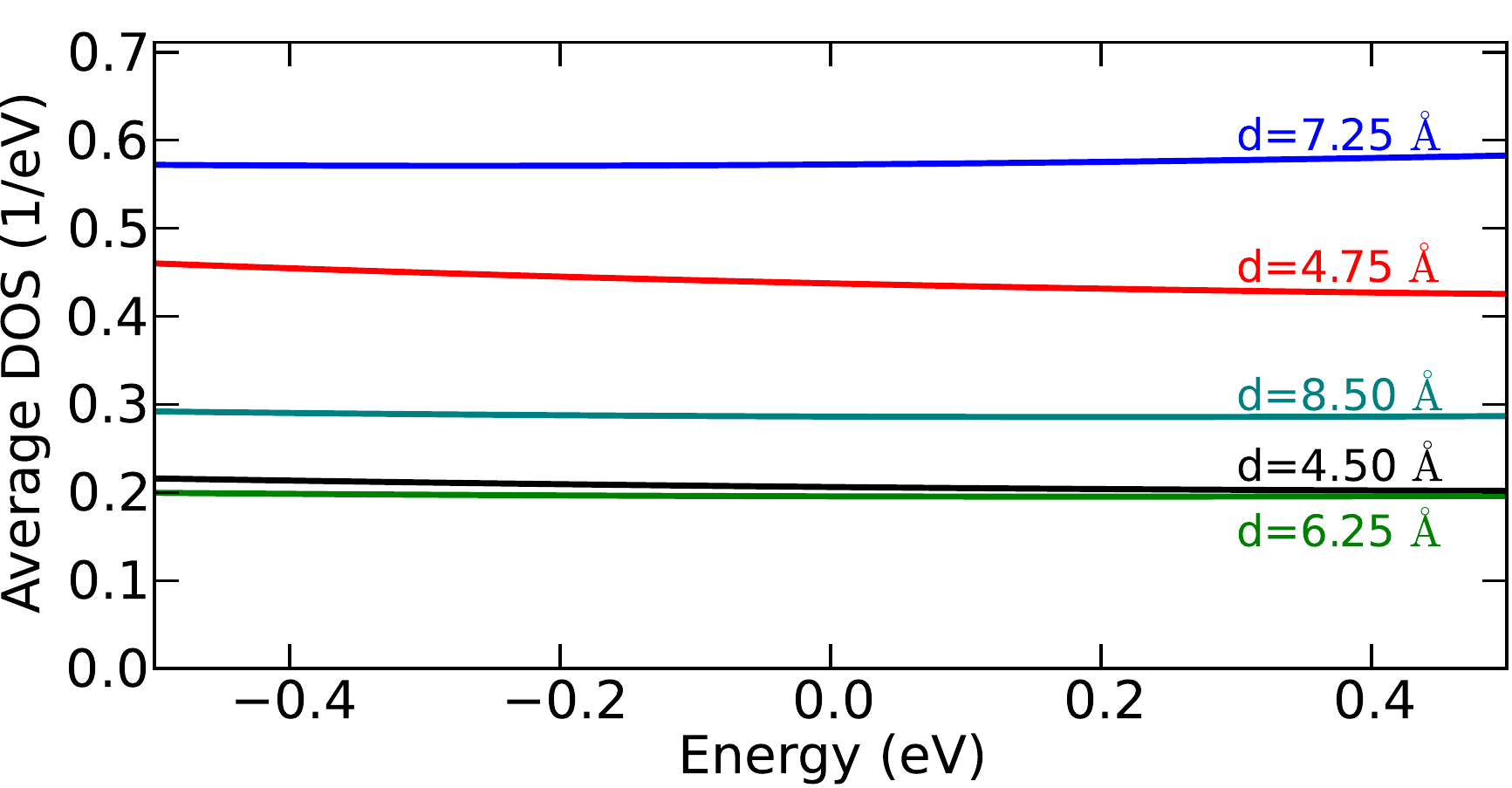}
\caption{\label{fig:distributions}(Color online) Top: Normalized distributions of current with the current axis on a log scale. The solid lines reflect the normalized distributions of current at pore diameters of 6.25, 7.25, and 8.5 {\AA}. The dashed lines are Gaussian fits to each distribution of $\mbox{log}\,I$. Bottom: Time averaged DOS as a function of energy referenced at the Fermi level of gold for different pore diameters.}
\end{figure}

In our analysis we cut out the first 1000 snapshots to eliminate any transient behavior. The currents from the remaining 4001 snapshots are binned to give a current distribution for each nanopore diameter, as in Fig. \ref{fig:distributions}. The distributions take the form of approximate Gaussian distributions, however the current axis is on a log scale. Therefore the distributions are approximately of the form

\begin{equation}
P(I) \sim \mbox{exp}\left\{\frac{-(\mbox{log}(I/\mu) + 3cs^2/2)^2}{2s^2}\right\},
\label{eq:gaussian}
\end{equation}

\noindent where $P(I)$ is the probability of realizing the current value $I$, $\mu$ is the average current, $c=\mbox{ln}\,10$, and $s$ is the standard deviation of the distribution of $\mbox{log}\,I$. This is equivalent to stating that the distribution of $\mbox{log}\,I$ is approximately a normal distribution. These observations suggest that the coupling between the electrodes and the water molecules is controlling the current distributions \cite{Krems2009}.

\begin{figure}
\includegraphics[width=\columnwidth]{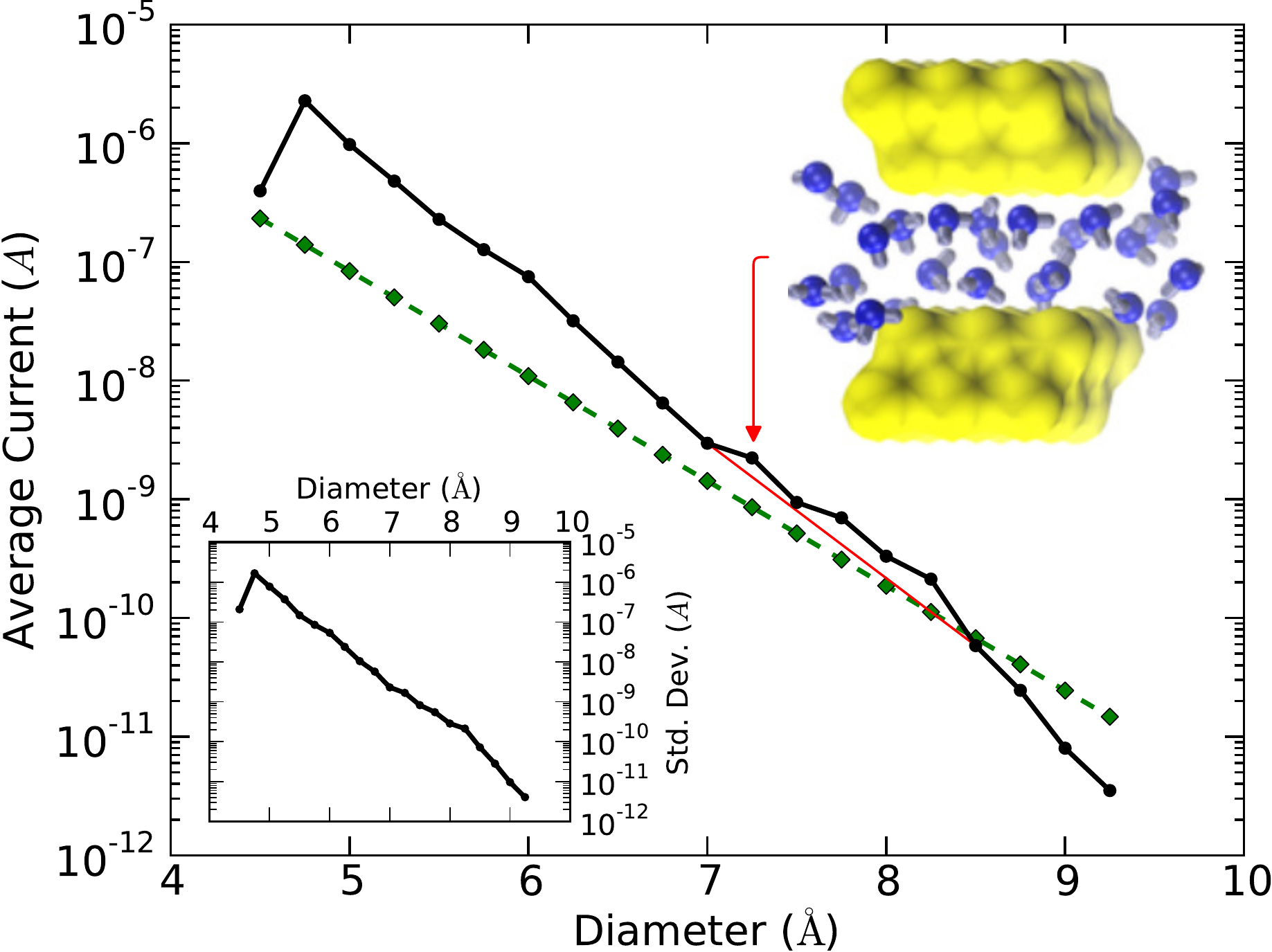}
\caption{\label{fig:length}(Color online) Average current plotted against pore diameter with the current axis on a log scale. The solid line reflects the time averaged current from the MD simulations while the dashed line reflects the current of a rectangular barrier (see text). The red line connects the points at 7 and 8.5 {\AA} and is there only to highlight the bump in the current. The lower inset demonstrates the standard deviation of the current from MD against pore diameter again with the current axis on a log scale. The upper inset, created with VMD \cite{HUMP96}, shows a snapshot of the gold (yellow) electrodes and the surrounding water (blue) at a pore diameter of 7.25 {\AA}.}
\end{figure}

The current averages are plotted against pore diameter in Fig. \ref{fig:length}. A simulation was run for every diameter from 4.5 to 9.25 {\AA} in 0.25 {\AA} intervals. Below 4.5 {\AA} water is completely excluded from the region between the electrodes and above 9.25 {\AA} we obtain sub-pA average currents which cannot be easily detected with present techniques. As a medium water acts to reduce the effective barrier height to about 1 eV \cite{Prokopuk2007, Albrecht2012}, much lower than the work function of gold (4.3 eV \cite{Ashcroft1976}). To emphasize this point we have calculated the current of a rectangular tunneling barrier in which the barrier height is the work function of gold and the barrier width is the diameter minus twice the distance between the edge of a jellium electron model with the gold density ($r_s=3$) and the center of the closest plane of gold atoms \cite{PhysRevB.68.157301}. The result of this calculation (dashed line in Fig. \ref{fig:length}) gives currents that are generally smaller than those of the MD simulations (solid line) until about 8.5 {\AA}. This is due to the simplistic choice of geometry of the barrier which becomes more influential as the pore diameter increases. The lower inset of Fig. \ref{fig:length} shows how the current standard deviation follows the same trend as the current average and generally decreases with increased pore diameter. Even so, the standard deviations and averages still differ and resemble each other because of the nature of the distributions of current in Eq. \ref{eq:gaussian} and the range of current values spanning one to two orders of magnitude as seen in Fig. \ref{fig:distributions}.

The first feature to notice in Fig. \ref{fig:length} is the deviation from the line of exponential dependence that includes diameters 7.25 to 8.25 {\AA}. Since the current axis is on a log scale, this deviation appears deceptively small. However the MD current values can be several times larger than the currents from a regression in the domain of the deviation. We now show that this increase in the tunneling current is the result of {\it structural changes} in the water.

\begin{figure}
\includegraphics[width=0.938\columnwidth]{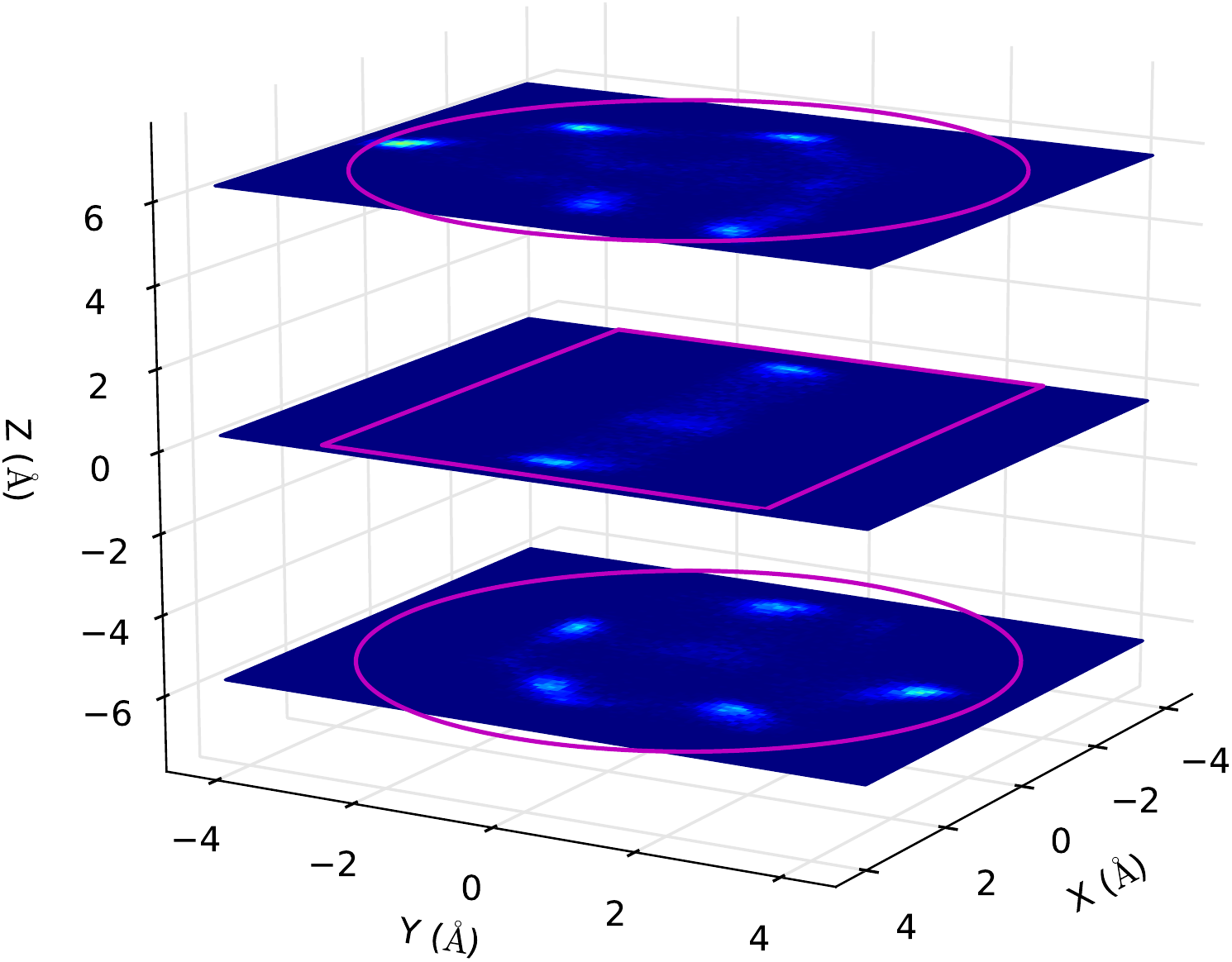}
\includegraphics[width=0.938\columnwidth]{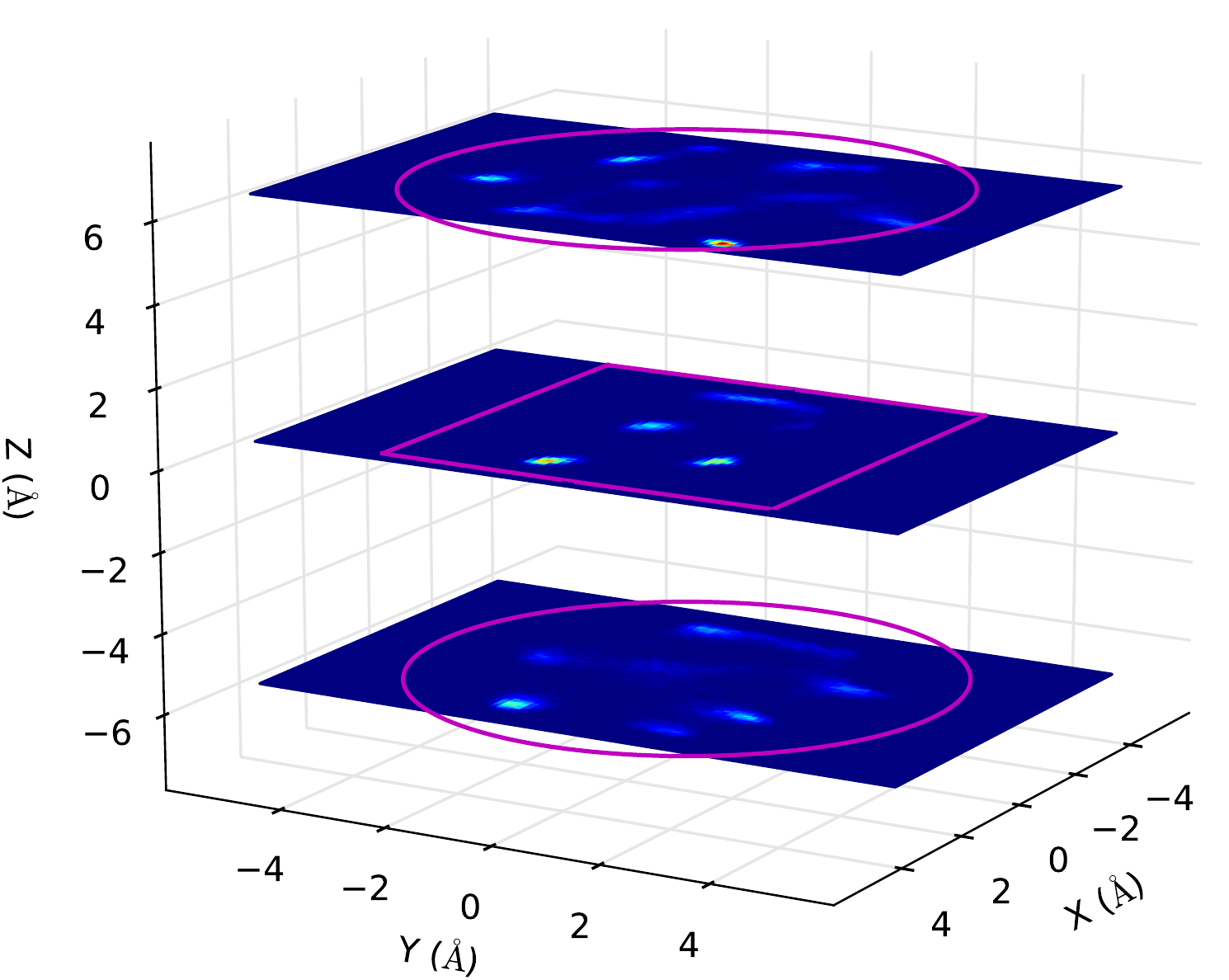}
\includegraphics[width=0.938\columnwidth]{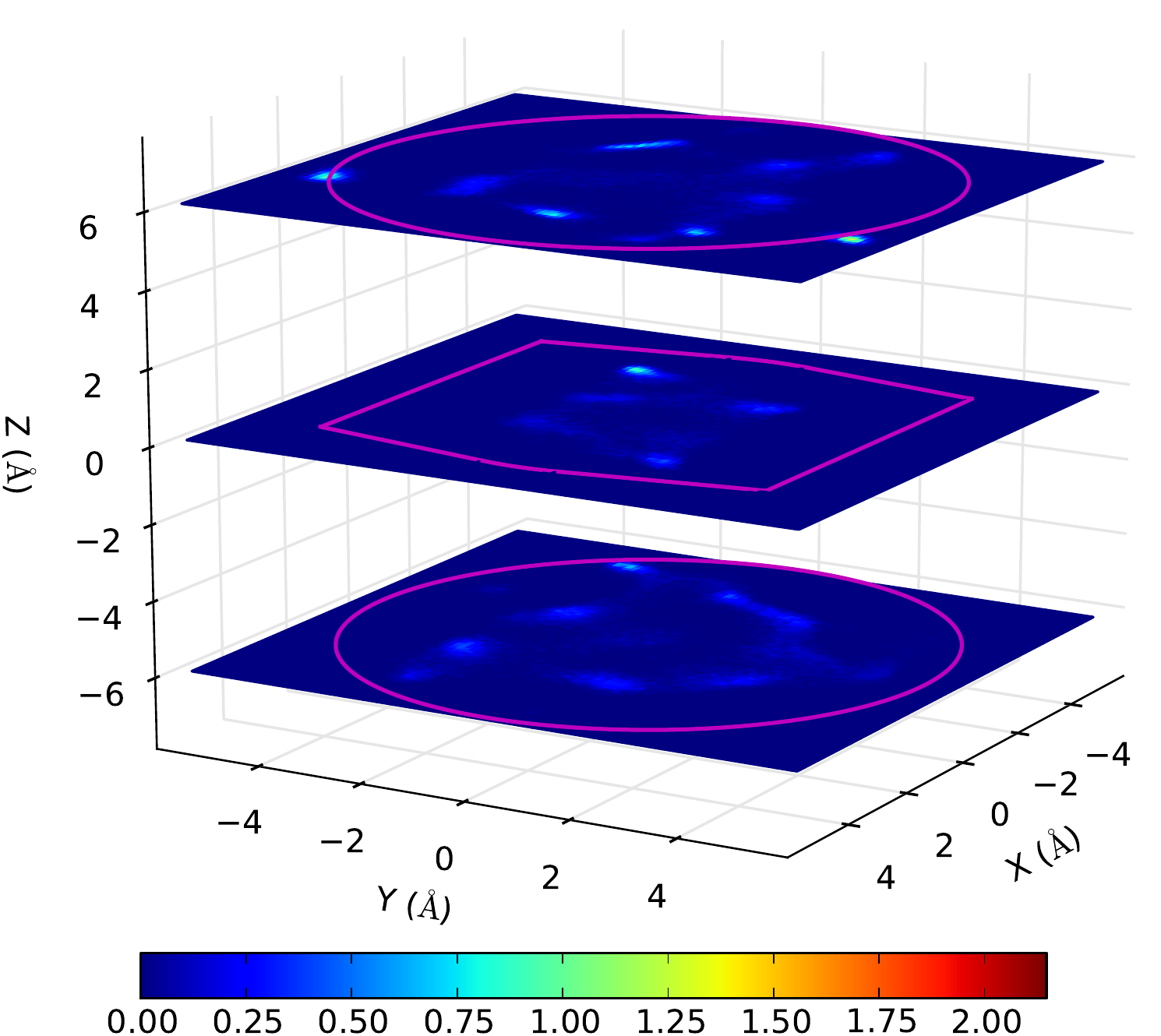}
\caption{\label{fig:wdp_high}(Color online) Time averaged water density plots for the (Top: 6.25, Center: 7.25, Bottom: 8.5) {\AA} diameter pore. The oxygen density ($\mbox{number}/\mbox{\AA}^3$) is displayed via color. The y-axis is the transverse electrode axis, the z-axis is the longitudinal axis, and the x-axis is the remaining transverse axis. The z-axis has been divided into five parts with the bottom, center, and top fifths of oxygen being shown. The magenta outlines represent the pore-electrode boundary.}
\end{figure}

To study the structure of water we plotted time averaged density profiles of oxygen for several pores as detailed in Fig. \ref{fig:wdp_high}. The number of water molecules in each snapshot can go from about 10 in the case of the 4.5 {\AA} pore diameter to about 45 in the case of 9.25 {\AA}. The density maps are plotted at the midpoint of each respective $z$-slice of space along the $(x,y)$ plane with the magenta outlines representing the pore-electrode boundary at the midpoint of each $z$-slice. At the center of the pore the boundary is dominated by the flat electrodes yielding near rectangular confinement, whereas the density above and below the electrodes shows the circular structures created in a cylindrical confinement.

We first notice that at 6.25 {\AA} only one layer of water can fit between the electrodes, although there is enough space for large fluctuations as seen by the blurred density. At about 7.25 {\AA} we see the formation of two layers of water packed together tightly. In fact, these layers start forming ``nanodroplets'' to reduce energy (see inset in Fig. \ref{fig:length}). Lastly at 8.5 {\AA} the bilayer of water is smeared all over the confined space implying large fluctuations again. These structural tendencies coordinated with length reinforce the conjecture that the deviation of the current from a simple exponential form is correlated to a sort of structural criticality. However, we have checked that these critical structures do not resemble a solid form of water. Instead thermal fluctuations cause water molecules to be exchanged with the bulk so that several molecules may explore the available space between the electrodes for an energetically favorable position.

In order to have a better understanding of the effect of these structural motifs we have computed the density of states (DOS) at the different diameters. In fact, the DOS roughly follows the same trend as that of the current (see bottom panel of Fig. \ref{fig:distributions}): at about 7.25 {\AA} the DOS is larger than at any other diameter. This implies that certain structural forms of water introduce more states for the tunneling electrons to utilize, effectively reducing the barrier between the electrodes, thus increasing the current. In the case of 6.25 {\AA} the DOS is the smallest of those shown in Fig. \ref{fig:distributions}, meaning that the fluctuating single layer of water that we see in Fig. \ref{fig:wdp_high} introduces less states for the tunneling electrons to utilize.

The last feature to notice in Fig. \ref{fig:length} is the abrupt change in current from 4.5 to 4.75 {\AA}. This corresponds to the first occasion in which water molecules can enter the space between the electrodes. At a pore diameter of 4.5 {\AA} the DOS remains low due to the exclusion of water, but increases dramatically when water molecules are confined between the electrodes at a pore diameter of 4.75 {\AA} (see bottom panel of Fig. \ref{fig:distributions}). Although the distance between the electrodes slightly increases, the tunneling current actually increases because of the introduction of water molecules and therefore an increased DOS.

In conclusion, we have shown with a combination of MD and quantum transport simulations the potential for probing the structure of confined water in nanopores with tunneling currents. We found that the distributions of the log of the current are normal, suggesting that the coupling between the electrodes and the water molecules governs the form of the distributions \cite{Krems2009}. We also find a highly non-linear dependence of the log of the current as a function of pore diameter. This non-linearity is due to the introduction of states for electrons to tunnel through when water molecules form nanodroplets between the electrodes. Because the effective tunneling barrier is reduced when electrons tunnel through water compared to vacuum \cite{Prokopuk2007, Albrecht2012}, we record currents in the range of pA to $\mu$A. These values as well as the recent demonstration of nanopores with embedded electrodes make our predictions within reach of experimental verification.

This work was supported by the National Institute of Health. We thank Chun-Keung Loong for useful discussions.
%
%
\bibliography{all_bibs}
\end{document}